\documentclass{cernyrep}
\usepackage[T1]{fontenc}
\usepackage[bookmarks, colorlinks=true, linktoc=page, linkcolor=black, citecolor=black, urlcolor=blue]{hyperref}
\newcommand{\der}[2]{\frac{\mathrm{d}#1}{\mathrm{d}#2}}
\newcommand{\pder}[2]{\frac{\partial#1}{\partial#2}}
\newcommand{\rd}[0]{\mathrm{d}}

\usepackage{fancyhdr}
\fancyhfoffset{4 mm}
\fancypagestyle{ARTTITLE}{%
\fancyhf{} 
\lhead{\small{Proceedings of the 2018 CERN--Accelerator--School course on \it{Beam Instrumentation}, Tuusula, (Finland)}} 
\lfoot{Available online at \url{https://cas.web.cern.ch/previous-schools}}
\rfoot{\thepage\hspace*{3mm}}
 
}

\usepackage{varwidth}
\usepackage{xcolor}

\begin{document}
\title{Longitudinal Beam Dynamics---Recap}
\author{F. Tecker}
\institute{CERN, Geneva, Switzerland}

\begin{abstract}
This paper gives a very brief summary of longitudinal beam dynamics for both linear and circular accelerators. After discussing synchronism conditions in linacs, it focuses on particle motion in synchrotrons. It summarizes the equations of motion, discusses phase-space matching during beam transfer, and introduces the Hamiltonian of longitudinal motion.
\end{abstract}
\maketitle
\thispagestyle{ARTTITLE}

\section{Introduction}
The force $\vec{F}$ on a charged particle with a charge $e$ is given by the Newton--Lorentz force:
\begin{equation}
\vec{F}=\frac{\mathrm{d}\vec{p}}{\mathrm{d}t} = e \left( \vec{E} + \vec{v} \times \vec{B} \right) \, .
\label{eq:force}
\end{equation}

The second term on the right-hand side is always perpendicular to the direction of motion, so it does not give any longitudinal acceleration and does not increase the energy of the particle.
Hence, the acceleration has to come from an electric field $\vec{E}$. To accelerate the particle, the field needs to have a component in the direction of motion of the particle. If we assume the field and the acceleration to be along the $z$ direction, Eq.~(\ref{eq:force}) becomes
\begin{equation}
\der{p}{t} = e E_z \, .
\end{equation}

The total energy $E$ of a particle is the sum of the rest energy $E_0$ and the kinetic energy $W$:
\begin{equation}
E = E_0 + W \, .
\end{equation}
In relativistic dynamics, the total energy $E$ and the momentum $p$ are linked by
\begin{equation}
E^2 = E_0^2 + p^2 c^2
\label{eq:E}
\end{equation}
(with $c$ being the speed of light), from which it follows that
\begin{equation}
\mathrm{d}E = v \, \mathrm{d}p \: .
\end{equation}

The rate of energy gain per unit length of acceleration (along the $z$ direction) is then given by
\begin{equation}
\der{E}{z} = v \der{p}{z} = \der{p}{t} = e E_z\, , 
\end{equation}
and the (kinetic) energy gained from the field along the $z$ path follows from $ \mathrm{d}W = \mathrm{d}E = e E_z \,\mathrm{d}z$:
\begin{equation}
W = e \int\! E_z \,\mathrm{d}z = eV \, ,
\end{equation}
where $V$ is just an electric potential.

The accelerating system will depend on the evolution of the particle velocity, which depends strongly on the type of particle.
The velocity is given by
\begin{equation}
v = \beta c = c \,\sqrt{1-\frac{1}{\gamma^2}} \; ,
\end{equation}
with the relativistic gamma factor of $\gamma = E / E_0$, the total energy $E$ divided by the rest energy $E_0$.
Electrons reach a constant velocity (close to the speed of light) at relatively low energies of a few mega\-electronvolts,
whereas heavy particles reach a constant velocity only at very high energies.
As a con\-sequence, one needs different types of resonator, optimized for different velocities.
In particular, this requires an acceleration system that remains synchronized with the particles during their acceleration.
For instance, when the revolution frequency in a synchrotron varies, the radio frequency (RF) will have to change correspondingly.

\section{Phase stability and energy-phase oscillation in a linac}

\begin{figure}[!h]
 \centering\includegraphics[width=0.85\textwidth]{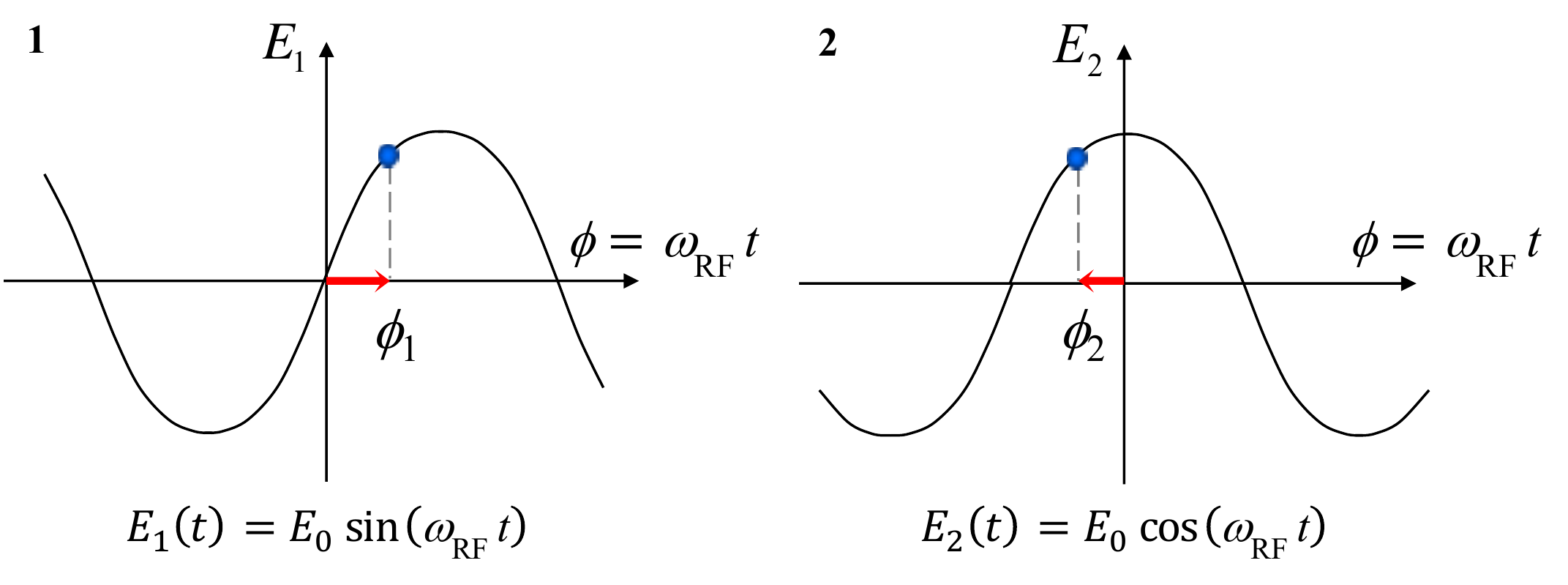}
 \caption{\label{fig:phase-conventions} Common phase conventions: (1)~the origin of time is taken at the zero-crossing with positive slope; (2)~the origin of time is taken at the positive crest of the RF voltage.}
\end{figure}

Several phase conventions exist in the literature (see Fig. \ref{fig:phase-conventions}):
\begin{itemize}
\item mainly for circular accelerators, the origin of time is taken at the zero-crossing with positive slope;
\item mainly for linear accelerators, the origin of time is taken at the positive crest of the RF voltage.
\end{itemize}
In the following, I will stick to the former convention of the positive zero-crossing for both the linear and the circular cases.

\begin{figure}
 \centering\includegraphics[width=0.95\textwidth]{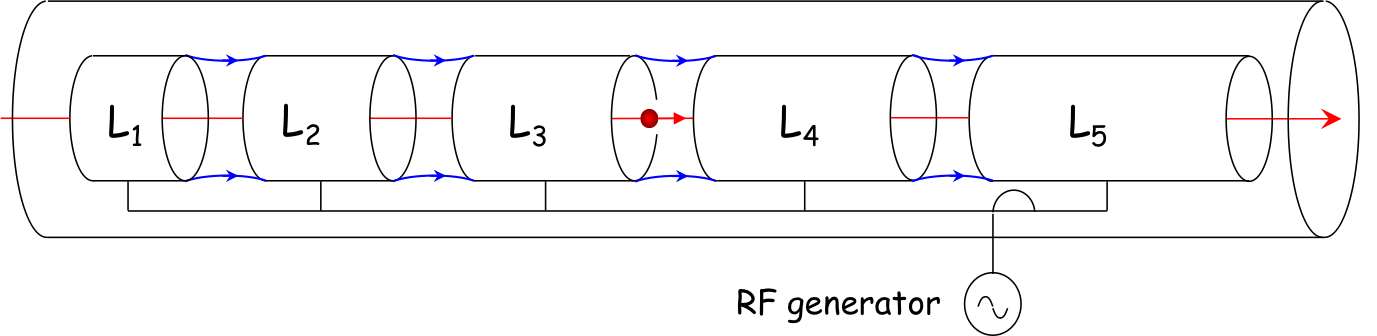}
 \caption{\label{fig:Alvarez} Alvarez-type accelerating structure}
\end{figure}

Let us consider an Alvarez structure, a succession of drift tubes in a single resonant tank connected to an RF oscillator that changes the potential of the tubes (see Fig.~\ref{fig:Alvarez}). A particle arriving in a gap during the time of an accelerating field will gain energy and be further accelerated.
Inside the tubes, the particle is shielded from the outside field. If the polarity of the field is identical in the next gap, the particle is again accelerated in that gap.
This leads to the synchronism condition that the distance $L$ between the gaps must satisfy
\begin{equation}
L = v T \, ,
\end{equation}
where $v = \beta \, c$ is the particle velocity and $T$ the period of the RF oscillator.
As the particle velocity increases, the drift spaces must get longer.
It is clear that this arrangement cannot accelerate a continuous beam; only a certain phase range will be accelerated and the beam must be bunched.

\begin{figure}[!b]
 \centering\includegraphics[width=0.9\textwidth]{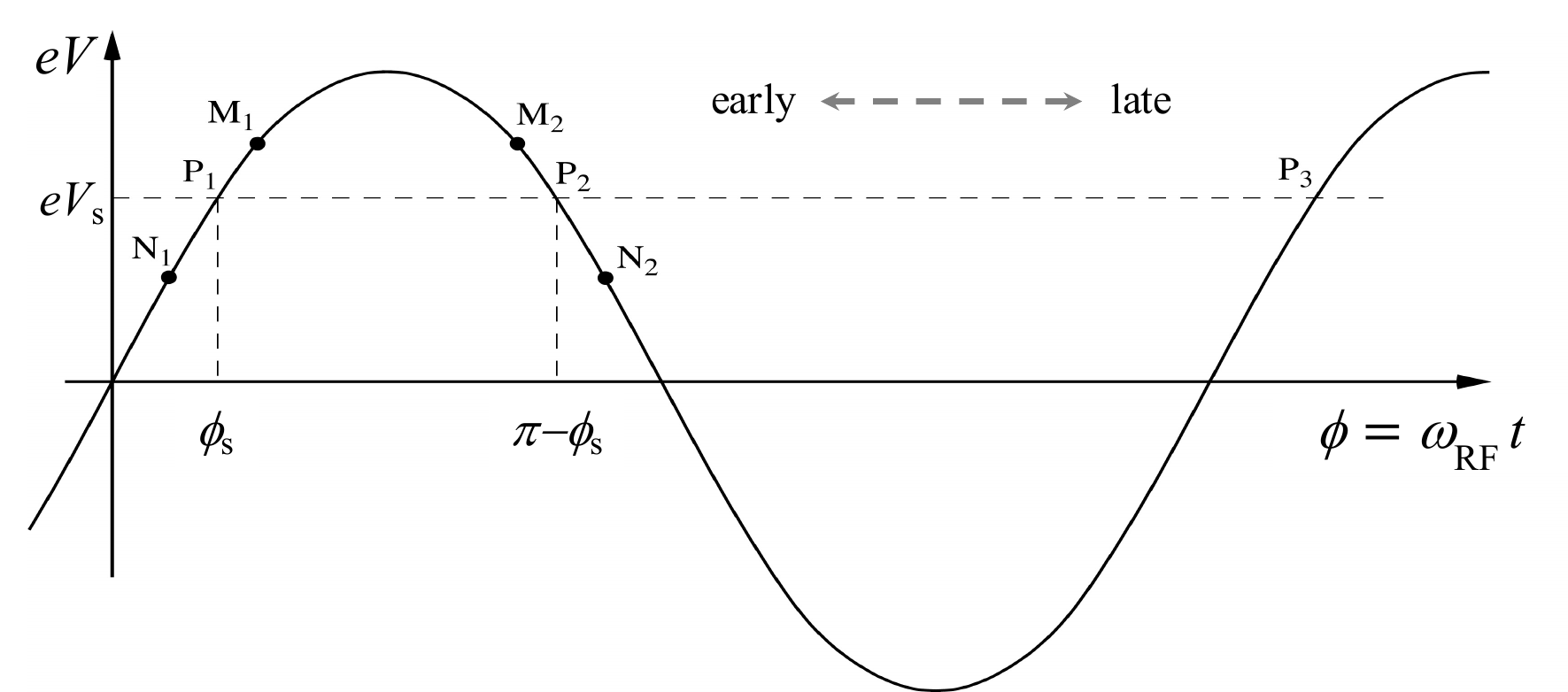}
 \caption{\label{fig:stability} Energy gain as a function of particle phase}
\end{figure}

By design, the energy gain for a particle passing through the structure at a certain RF phase $\phi_{\rm s}$ is such that the particle reaches the next gap with the same phase $\phi_{\rm s}$. Then the energy gain in the following gap will again be the same, and the particle will pass all gaps at this phase $\phi_{\rm s}$, which is called the `synchronous phase'. So the energy gain is $e V_{\rm s} = e \hat{V} \sin \phi_{\rm s}$. This is illustrated in Fig.~\ref{fig:stability} by the points P$_1$ and P$_3$.

A particle N$_1$ that arrives in a gap earlier than P$_1$ will gain less energy and have smaller
 velocity, so it will take more time to travel through the drift tube. In the
next gap, it will appear closer to particle P$_1$. The effect is opposite for particle M$_1$ in Fig.~\ref{fig:stability}, which will gain
more energy and reduce its delay compared with P$_1$. So the points P$_1$, P$_3$, \etc are stable points
for the acceleration, since particles pos\-itioned slightly away from them will experience forces that 
reduce their deviation. In contrast, it can be seen that point P$_2$ is an unstable point,
in the sense that particles slightly away from it will deviate even more in the next gaps.

Thus, for stability of the longitudinal oscillation, the particle needs to be on the rising slope of the RF field to have a restoring force towards the stable phase.

To study longitudinal motion, it is convenient to use  variables that represent the phase
and energy relative to the synchronous particle (denoted by the subscript `s'):
\begin{align}
\varphi & =  \phi - \phi_{\rm s} \, , \\
w & =  E - E_{\rm s} = W - W_{\rm s} \,.
\end{align}
The accelerating field can be described simply by
\begin{equation}
E_z = E_0 \sin(\omega t) \, .
\end{equation}

The rate of energy gain for the synchronous particle is 
\begin{equation}
\frac{\mathrm{d}E_{\rm s}}{\mathrm{d}z} = \der{p_{\rm s}}{t} = e E_0 \sin \phi_{\rm s}\: ,
\end{equation}
and that for a non-synchronous particle (for small $\varphi$) is
\begin{equation}
\der{w}{z} = e E_0 \bigl[ \sin(\phi_{\rm s} + \varphi) - \sin\phi_{\rm s} \bigr]  \approx e E_0 \cos\phi_{\rm s} \, \varphi \: .
\label{eq:w}
\end{equation}

The rate of change of the phase with respect to the synchronous particle is, for small deviations,
\begin{equation}
 \der{\varphi}{z} = \omega_{\rm RF} \left[ \der{t}{z} - \left( \!\der{t}{z}\! \right)_{\!\!{\rm s}} \,\right] =
 \omega_{\rm RF}  \left( \frac{1}{v} - \frac{1}{v_{\rm s}} \right) \approx - \frac{\omega_{\rm RF}}{v^2_{\rm s}} (v-v_{\rm s}) \: .
\end{equation}
Writing $\rd \gamma = \gamma^3 \beta \, \rd\beta$, $w$ becomes, in the vicinity of the synchronous particle,
\begin{equation}
w = E -E_{\rm s} = m_0 c^2 ( \gamma - \gamma_{\rm s} ) = m_0 c^2 \,\rd\gamma = m_0 c^2 \gamma_{\rm s}^3 \beta_{\rm s} \, \rd\beta = m_0 \gamma_{\rm s}^3 v_{\rm s} (v - v_{\rm s}) \, ,
\end{equation}
which leads to
\begin{equation}
 \der{\varphi}{z} = - \frac{\omega_{\rm RF}}{m_0 v_{\rm s}^3 \gamma_{\rm s}^3}\, w \, .
 \label{eq:phi}
\end{equation}

Combining the two first-order equations (Eqs.\ (\ref{eq:w}) and (\ref{eq:phi})) into a second-order equation gives the equation of a harmonic oscillator with angular frequency $\Omega_{\rm s}$:
\begin{equation}
\der{^2\varphi}{z^2} + \Omega_{\rm s}^2 \varphi = 0  \qquad\text{where }\,\Omega_{\rm s}^2 = \frac{eE_0\omega_{\rm RF}\cos\phi_{\rm s}}{m_0v_{\rm s}^3\gamma_{\rm s}^3} \, .
\label{eq:lin-osc}
\end{equation}
Stable harmonic oscillations imply that $\Omega_{\rm s}^2$ is real and positive, which means that $\cos\phi_{\rm s} > 0$. Since acceleration implies that $\sin\phi_{\rm s} >0$, it follows that the stable phase region for acceleration in the linac is
\begin{equation}
0 < \phi_{\rm s} < \frac{\pi}{2} \: ,
\end{equation}
which confirms what we have seen before in our discussion of the restoring force towards the stable phase.

From Eq. (\ref{eq:lin-osc}) it is also clear that the oscillation frequency decreases strongly with increasing vel\-ocity (and relativistic gamma factor) of the particle. For highly relativistic particles (such as high-energy electrons) the velocity change is negligible, so there is practically no change in the particle phase, and the bunch distribution is no longer changing.

\section{Synchrotron}
\begin{figure}
 \centering\includegraphics[width=0.55\textwidth]{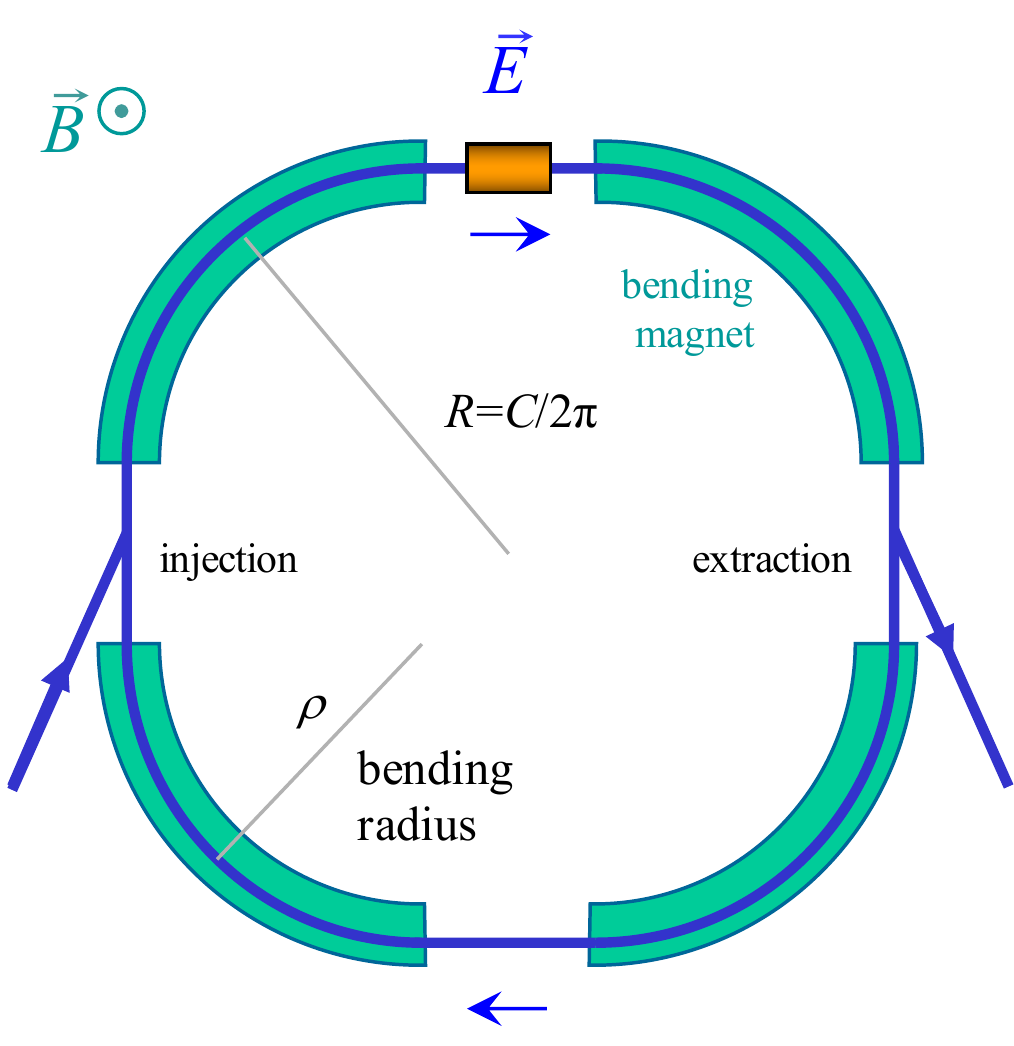}
 \caption{\label{fig:synchrotron} Synchrotron}
\end{figure}

A synchrotron (see Fig. \ref{fig:synchrotron}) is a circular accelerator in which the nominal particle trajectory is maintained at a
constant physical radius by varying both the magnetic field and the RF  to follow the energy variation.
In this way, the aperture of the vacuum chamber and the magnets can be kept small.

The RF needs to be synchronous to the revolution frequency. To achieve synchronism, the synchron\-ous particle needs to arrive at the cavity again after one turn with the same phase. This implies that the angular RF, $\omega_{\rm RF} = 2\pi f_{\rm RF}$, must be an integer multiple of the angular revolution frequency~$\omega$:
\begin{equation}
\omega_{\rm RF} = h \, \omega \, ,
\end{equation}
where $h$ is an integer called the {\it harmonic number}. As a consequence, the number of stable synchron\-ous particle locations equals the harmonic number $h$. These locations are equidistantly spaced around the circumference of the accelerator. All synchronous particles will have the same nominal energy and will follow the nominal trajectory.

Energy ramping is obtained by varying the magnetic field while following the change of the revolution frequency with a change in the RF.
The time derivative of the momentum,
\begin{equation}
p = e B \rho \, ,
\end{equation}
yields (when keeping the bending radius $\rho$ constant)
\begin{equation}
\der{p}{t} = e \rho \dot{B} \, .
\end{equation}
For one turn in the synchrotron, this results in
\begin{equation}
(\Delta p)_\mathrm{turn} = e \rho \dot{B} T_{\rm r} = \frac{2 \pi e \rho R \dot{B}}{v} \: ,
\end{equation}
where $R=L / (2\pi)$ is the physical radius of the machine.

Since $E^2 = E_0^2 + p^2 c^2$,
it follows that
$\Delta E = v \Delta p$, so 
\begin{equation}
(\Delta E)_\mathrm{turn} = (\Delta W)_{\rm s} = 2 \pi e \rho R \dot{B} = e \hat{V} \sin \phi_{\rm s} \, .
\end{equation}

From this relation it can be seen that the stable phase for the synchronous particle changes during the acceleration, when the magnetic field $B$ changes, as
\begin{equation}
\sin \phi_{\rm s} = 2 \pi \rho R \, \frac{\dot{B}}{\hat{V}_{\rm RF}} \qquad \mathrm{or} \qquad \phi_{\rm s} = \arcsin\left( 2 \pi \rho R \, \frac{\dot{B}}{\hat{V}_{\rm RF}} \right) \, .
\end{equation}

As mentioned previously, the RF has to follow the change of revolution frequency and will increase during acceleration as
\begin{equation}
\frac{f_{\rm RF}}{h} = f_{\rm r} = \frac{v(t)}{2\pi R_{\rm s}} = \frac{1}{2\pi}\, \frac{ec^2}{E_{\rm s}(t)} \,\frac{\rho}{R_{\rm s}}\, B(t) \, .
\end{equation}

Since $E^2 = E_0^2 + p^2 c^2$,
the RF must follow the variation of the $B$ field with the law
\begin{equation}
\frac{f_{\rm RF}}{h}  = \frac{c}{2\pi R_{\rm s}} \left\{ \frac{B(t)^2}{[m_0c^2 / (e c \rho)]^2 + B(t)^2} \right\}^{1/2} \, .
\end{equation}
This asymptotically approaches $f_{\rm r} \rightarrow {c}/({2\pi R_{\rm s}})$ as  $v \rightarrow c$ and $B$ becomes large compared with $m_0c^2 / (e c \rho)$.

\subsection{Dispersion effects in a synchrotron}
If a particle is slightly shifted in momentum, it will have a different velocity and also a different orbit and orbit length.
We can define two parameters: 
\begin{itemize}
\item the {\it momentum compaction factor} $\alpha_\mathrm{c}$, which is the relative change in orbit length with momentum, given by 
\begin{equation}
\alpha_\mathrm{c} = \frac{\Delta L/L}{\Delta p / p} \, ;
\end{equation}
\item the {\it slip factor} $\eta$, which is the relative change in revolution frequency with momentum, given by
\begin{equation}
\eta = \frac{\Delta f_{\rm r}/f_{\rm r}}{\Delta p / p} 
\end{equation}
(sometimes also defined in the literature with the opposite sign).
\end{itemize}

\begin{figure}
\begin{tabular}{p{0.4\textwidth} p{0.4\textwidth}}
  \vspace{0pt}\hspace{1cm} \includegraphics[width=0.3\textwidth]{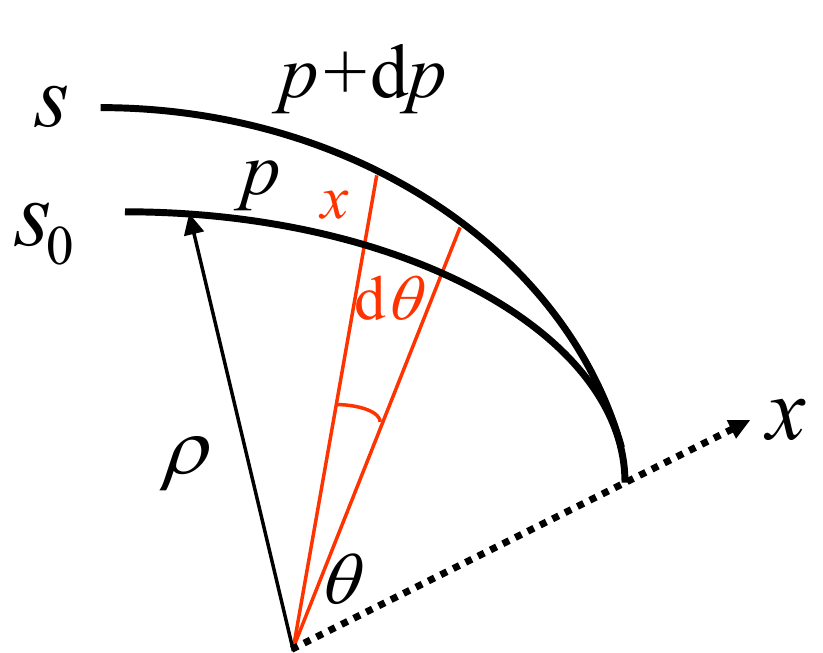} &
 \vspace{20pt}\begin{eqnarray}
\mathrm{d}s_0 & = & \rho \, \mathrm{d}\theta \, , \nonumber\\
\mathrm{d}s & = & ( \rho + x ) \, \mathrm{d}\theta \, . \nonumber
\end{eqnarray}
\end{tabular}
 \caption{\label{fig:orbit} Orbit length change}
\end{figure}

Let us consider the change in orbit length (see Fig. \ref{fig:orbit}). The relative elementary path-length difference for a particle with momentum $p + \mathrm{d}p$ is
\begin{equation}
\der{l}{s_0} = \der{s - \mathrm{d}s_0}{s_0} = \frac{x}{\rho} = \frac{D_x}{\rho} \frac{\mathrm{d}p}{p} \, ,
\end{equation}
where $D_x = \mathrm{d}x / (\mathrm{d}p/p)$ is the {\it dispersion function} from the transverse beam optics.

This leads to a total change in the circumference $L$ of
\begin{equation}
\mathrm{d}L = \int_C \mathrm{d}l = \int \frac{x}{\rho} \,\mathrm{d}s_0 = \int \frac{D_x}{\rho} \frac{\mathrm{d}p}{p} \,\mathrm{d}s_0 \, ,
\end{equation}
so that
\begin{equation}
\alpha_\mathrm{c} = \frac{1}{L} \int \frac{D_x}{\rho} \,\mathrm{d}s_0 \,
.
\end{equation}

Since $\rho = \infty$ in the straight sections, we get
\begin{equation}
\alpha_\mathrm{c} = \frac{\langle D_x \rangle_{\rm m}}{R} \, ,
\end{equation}
where the average $\langle \:\cdot\: \rangle_{\rm m}$ is considered over the bending magnets only.

Given that the revolution frequency is $f_{\rm r} = \beta c / (2\pi R)$, the relative change is (using the definition of the momentum compaction factor)
\begin{gather}
\frac{\rd f_{\rm r}}{f_{\rm r}} = \frac{\rd\beta}{\beta} - \frac{\rd R}{R} = \frac{\rd\beta}{\beta} - \alpha_\mathrm{c} \frac{\rd p}{p} \, , \\
p = mv = \beta \gamma \frac{E_0}{c} \;\Rightarrow \; \frac{\rd p}{p} =  \frac{\rd\beta}{\beta} + \frac{\rd(1-\beta^2)^{-1/2}}{(1-\beta^2)^{-1/2}} = \underbrace{\left( 1-\beta^2 \right)^{-1}}_{\displaystyle\gamma^2} \frac{\rd\beta}{\beta} \, .
\end{gather}

Hence, the relative change in revolution frequency is 
\begin{equation}
\frac{\rd f_{\rm r}}{f_{\rm r}} = \left( \frac{1}{\gamma^2} - \alpha_\mathrm{c} \right) \frac{\rd p}{p} \, ,
\end{equation}
which means that the slip factor $\eta$ is given by

\begin{equation}
\eta = \frac{1}{\gamma^2} - \alpha_\mathrm{c} \, .
\end{equation}

Obviously, there is one energy with a given $\gamma_{\rm t}$ for which $\eta$ becomes zero, meaning that there is no change in the revolution frequency for particles with a small momentum deviation. This energy is a property of the transverse lattice, with
\begin{equation}
\gamma_{\rm t} = \frac{1}{\sqrt{\alpha_\mathrm{c}}} \: .
\end{equation}

From the definition of $\eta$, it is clear that an increase in momentum gives
the following.\begin{itemize}
\item {\it Below transition} energy $(\eta > 0)$: a higher revolution frequency. The increase in the velocity of the particle is the dominating effect.
\item {\it Above transition} energy $(\eta < 0)$: a lower revolution frequency. The particle has a velocity close to the speed of light; this velocity does not change significantly any more. Thus, here the effect of the longer path-length dominates (for the most common case of transverse lattices with a positive momentum compaction factor,  $\alpha_\mathrm{c}>0$).
\end{itemize}

At transition, the velocity change and the path-length change with momentum compensate for each other,
so the revolution frequency there is independent of the momentum deviation.
As a consequence, the longitudinal oscillation stops and the particles in the bunch will not change their phase.
Particles that are not at the synchronous phase will get the same non-nominal energy gain in each turn and will accumulate an energy error that will increase the longitudinal emittance and can lead to a loss of the particle due to dispersive effects. Therefore, transition has to be passed quickly to minimize the emittance increase and the losses.

Electron synchrotrons do not need to cross transition. Owing to the relatively small rest mass of the electron, the relativistic gamma factor is so large
that  the injection energy is already greater than the transition energy. Hence, the electrons will stay above transition during the whole acceleration cycle.

\begin{figure}
 \centering\includegraphics[width=0.9\textwidth]{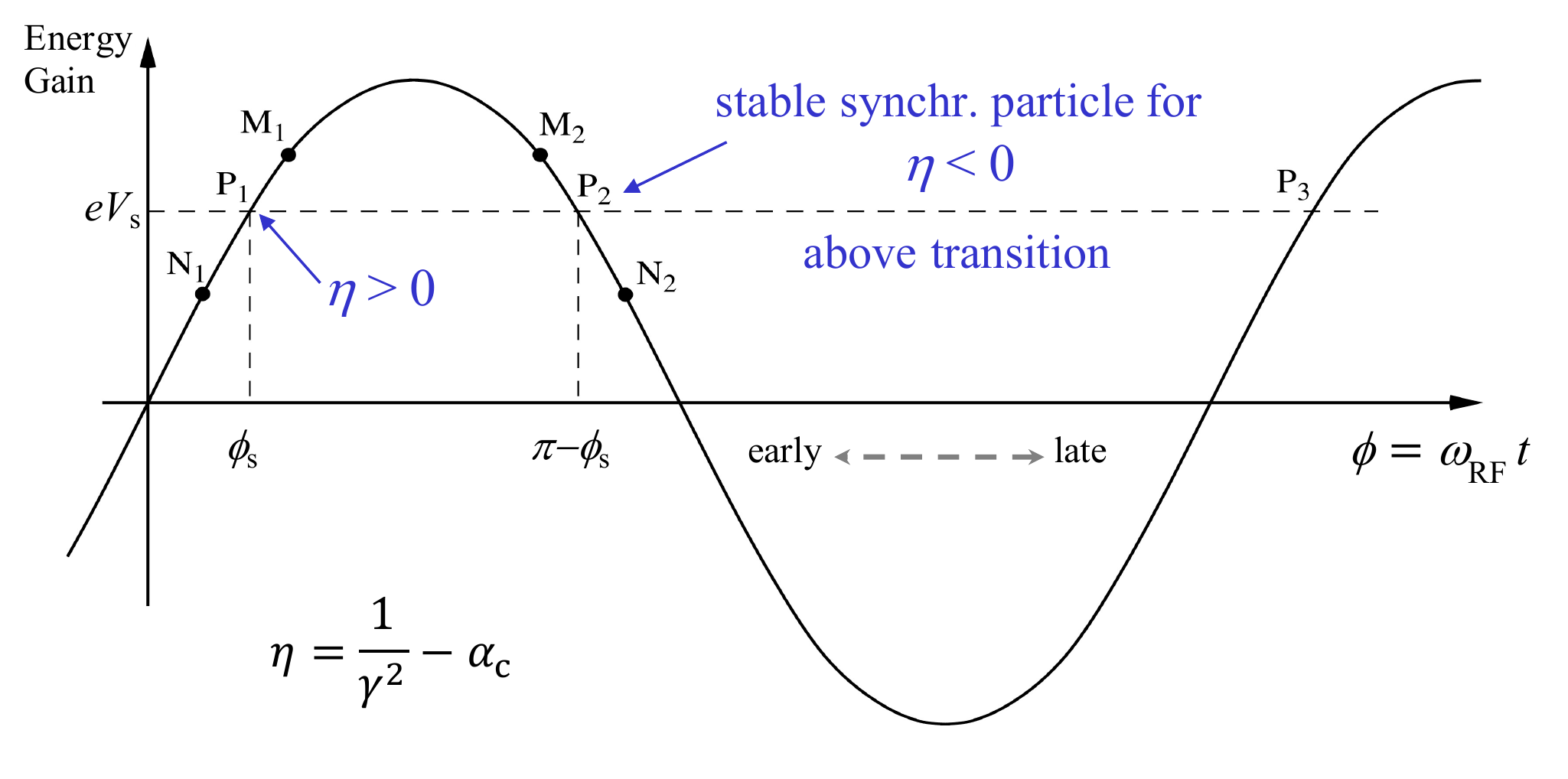}
 \caption{\label{fig:phase-stability} Energy gain as a function of particle phase; oscillations are stable around the synchronous-phase particle P$_1$ below transition and around the synchronous-phase particle P$_2$ above transition.}
\end{figure}

Since the changes in revolution frequency with momentum are opposite below and above transition, this completely alters the range for stable oscillations. As we have seen in the linac case (see Fig.~\ref{fig:stability}), the oscillation is stable for a particle that is on the rising slope of the RF field when we are below transition. Above transition, the oscillation is unstable and the stable region for oscillations is on the falling slope (see Fig.~\ref{fig:phase-stability}). A particle that arrives too early (M$_2$) will gain more energy, and the revolution time will increase owing to the predominant effect of the longer path; thus, the particle will arrive later on the next turn, closer to the synchronous phase. Similarly, a particle that arrives late (N$_2$) will gain less energy and travel a shorter orbit, also moving towards the synchronous phase.

\begin{figure}[!b]
 \centering\includegraphics[width=0.6\textwidth]{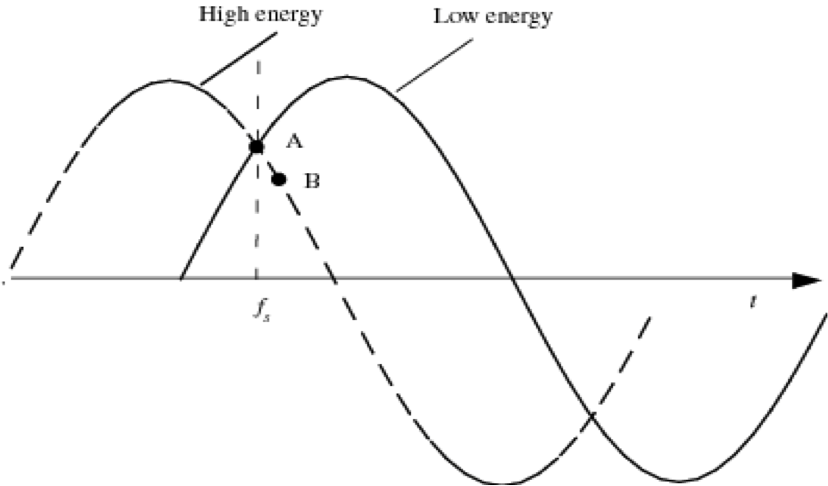}
 \caption{\label{fig:transition} The synchronous phase changes from a rising to a falling slope as the energy crosses transition~\cite{wilson}}
\end{figure}

Crossing transition during acceleration makes the previous stable synchronous phase unstable. Below transition, it is stable on the rising slope of the RF; above transition, the synchronous phase is on the falling slope. Consequently, the RF system needs to make a rapid change of RF phase when crossing the transition energy---a `phase jump', as indicated in Fig.~\ref{fig:transition}---otherwise the particles in the bunch become dispersed, have a wrong energy gain, and eventually get lost.
A method to improve transition crossing is to change the transverse optics when the energy almost reaches $\gamma_{\rm t}$ for optics with a larger momentum compaction factor $\alpha_\mathrm{c}$. Hence, $\gamma_{\rm t}$ is decreasing at the same time as the energy is increasing, and so the time with an energy close to transition can be reduced.

\subsection{Equations of longitudinal motion in a synchrotron}
\begin{figure}
\begin{center}
\begin{minipage}[t]{0.3\textwidth}
\vspace*{10pt}\includegraphics[width=0.7\textwidth]{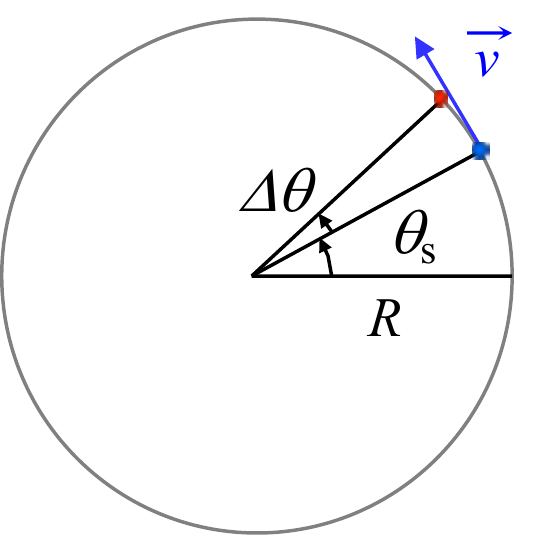}
\end{minipage}\hspace*{-5mm}
\begin{minipage}[t]{0.49\textwidth}
\vspace*{30pt}\begin{tabular}{ll}
particle RF phase & $\Delta\phi = \phi - \phi_{\rm s}\:$, \\
particle momentum & $\Delta p = p - p_{\rm s}\:$, \\
particle energy & $\Delta E = E - E_{\rm s}\:$, \\
azimuth angle& $\Delta\theta = \theta - \theta_{\rm s}\: .$
\end{tabular}
\end{minipage}
\end{center}
\caption{Variables with respect to synchronous particle}
\label{fig:Fred}
\end{figure}

As previously done for the linac, we want to look at the oscillations with respect to the synchronous particle, and we express the variables with respect to the synchronous particle, as shown in Fig.~\ref{fig:Fred}.

Since the RF is a multiple of the revolution frequency, the RF phase $\Delta\phi$ changes as
\begin{equation}
f_\mathit{\rm RF} = h\, f_{\rm r} \quad \Longrightarrow \quad \Delta\phi = -h \, \Delta\theta \quad \text{with }\: \theta = \int \omega_{\rm r}\, \rd t \, .
\end{equation}
The minus sign for the RF phase arises from the fact that a particle that is ahead arrives earlier, i.e.\ at a smaller RF phase.

For a given particle with respect to the reference particle, the change in angular revolution frequency is
\begin{equation}
\Delta\omega_{\rm r} = \der{}{t} \left( \Delta\theta \right) = - \frac{1}{h} \der{}{t} \left( \Delta\phi \right) = - \frac{1}{h} \der{\phi}{t} \: .
\end{equation}

Since
 \[
 \displaystyle\eta = \frac{p_{\rm s}}{\omega_{\rm rs}} \left( {\frac{\mathrm{d}\omega_{\rm r}}{\mathrm{dp}}} \right)_{\!{\rm s}} \, ,
\]
 $E^2 = E_0^2 + p^2 c^2$, and $\Delta E = v_{\rm s} \Delta p = \omega_{\rm rs} R_{\rm s} \Delta p$, one gets the first-order equation
\begin{equation}
\frac{\Delta E}{\omega_{\rm rs}} = - \frac{p_{\rm s} R_{\rm s}}{h \eta \omega_{\rm rs}} \der{(\Delta\phi)}{t} = - \frac{p_{\rm s} R_{\rm s}}{h \eta \omega_{\rm rs}} \, \dot{\phi} \, . \label{eq:first1}
\end{equation}
The second first-order equation follows from the energy gain of a particle:
\begin{align}
\der{E}{t} &= \frac{\omega_{\rm r}}{2\pi} \,e \hat{V} \sin\phi \, , \\
2 \pi \der{}{t} \left( \frac{\Delta E}{\omega_{\rm rs}} \right) &= e \hat{V} \bigl( \sin\phi - \sin\phi_{\rm s} \bigr)\: . \label{eq:first2}
\end{align}

Combining the two first-order equations (Eqs. (\ref{eq:first1}) and (\ref{eq:first2})) leads to
\begin{equation}
\der{}{t} \left[ \frac{R_{\rm s} p_{\rm s}}{h \eta \omega_{\rm rs}} \der{\phi}{t} \right] + \frac{e \hat{V}}{2\pi}  ( \sin\phi - \sin\phi_{\rm s}) = 0 \, .
\label{eq:gen2}
\end{equation}
This second-order equation is non-linear, and the parameters within the bracket are, in general, slowly varying in time.

When we assume constant parameters $R_{\rm s}, p_{\rm s}, \omega_{\rm s}$, and $\eta$, we get
\begin{equation}
\ddot\phi + \frac{\Omega_{\rm s}^2}{\cos\phi_{\rm s}}  \left( \sin\phi - \sin\phi_{\rm s} \right) = 0 \qquad \text{with } \: \Omega_{\rm s}^2 = \frac{h \eta \omega_{\rm rs} e \hat{V} \cos\phi_{\rm s}}{2\pi R_{\rm s} p_{\rm s}} \label{eq:osc}
\end{equation}
and, for small phase deviations from the synchronous particle,
\begin{equation}
\sin\phi - \sin\phi_{\rm s} = \sin(\phi_{\rm s}+\Delta\phi) - \sin\phi_{\rm s} \approxeq \cos\phi_{\rm s} \Delta\phi \, ,
\end{equation}
so that we end up with the equation of a harmonic oscillator:
\begin{equation}
\ddot\phi + \Omega_{\rm s}^2 \Delta\phi = 0 \, ,
\end{equation}
where $\Omega_{\rm s}$ is the synchrotron angular frequency.

\begin{figure}[h]
 \centering\includegraphics[width=0.95\textwidth]{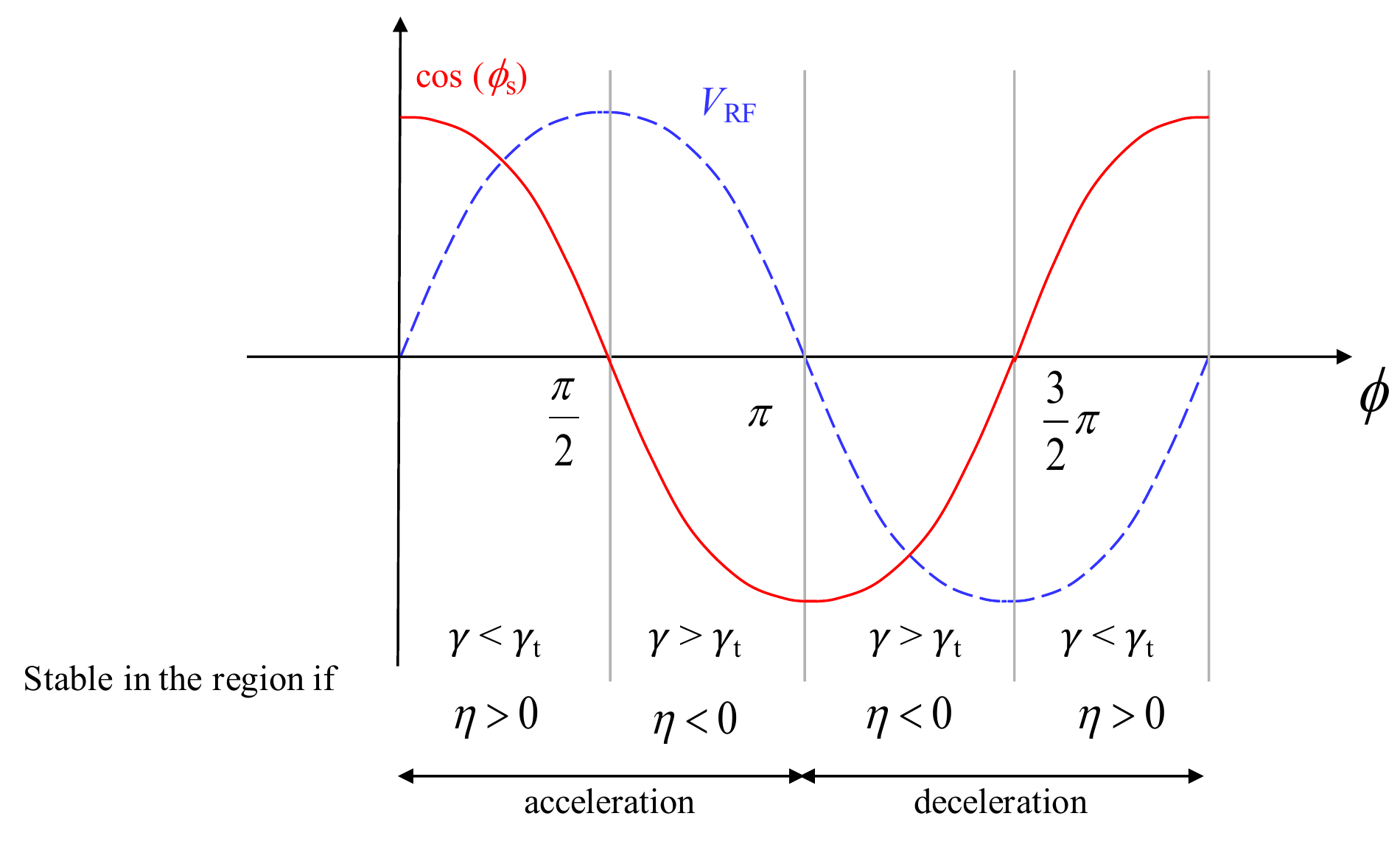}
 \caption{\label{fig:stability-regions} Stability regions as a function of particle phase, depending on the energy with respect to transition
}
\end{figure}

Stability is obtained when $\Omega_{\rm s}$ is real so that $\Omega_{\rm s}^2$ is positive. Since most terms in the expression for $\Omega_{\rm s}^2$ are positive, this condition reduces to
\begin{equation}
\eta \cos\phi_{\rm s} > 0\: ,
\end{equation}
and the stable region for the synchronous phase depends on the energy relative to the transition energy, as we have seen from our previous argument. The conditions for stability are summarized in Fig.~\ref{fig:stability-regions}.

The \emph{synchrotron tune} $Q_{\rm s}$ is defined as
\begin{equation}
Q_{\rm s} = \Omega_{\rm s} / \omega_{\rm r} 
\end{equation}
and corresponds to the number of synchrotron oscillations per revolution in the synchrotron. Generally $Q_{\rm s} \ll 1$, as it  typically
takes of the order of several hundreds of turns to complete one synchrotron oscillation.

For larger phase (or energy) deviations from the synchronous particle, we can multiply Eq. (\ref{eq:osc}) by $\dot\phi$ and integrate it, obtaining an invariant of motion
\begin{equation}
\frac{\dot\phi^2}{2} - \frac{\Omega_{\rm s}^2}{\cos\phi_{\rm s}}  \left( \cos\phi + \phi \sin\phi_{\rm s} \right) = I \, ,
\label{eq:invariant}
\end{equation}
which for small amplitudes of $\Delta\phi$ reduces to
\begin{equation}
\frac{\dot\phi^2}{2} + \Omega_{\rm s}^2 \,\frac{(\Delta\phi)^2}{2} = I' \,
.
\end{equation}
Similar equations can be derived for the second variable $\Delta E \propto \mathrm{d}\phi
/ \mathrm{d}{t}$.

\begin{figure}[htb]
 \centering\includegraphics[width=0.6\textwidth]{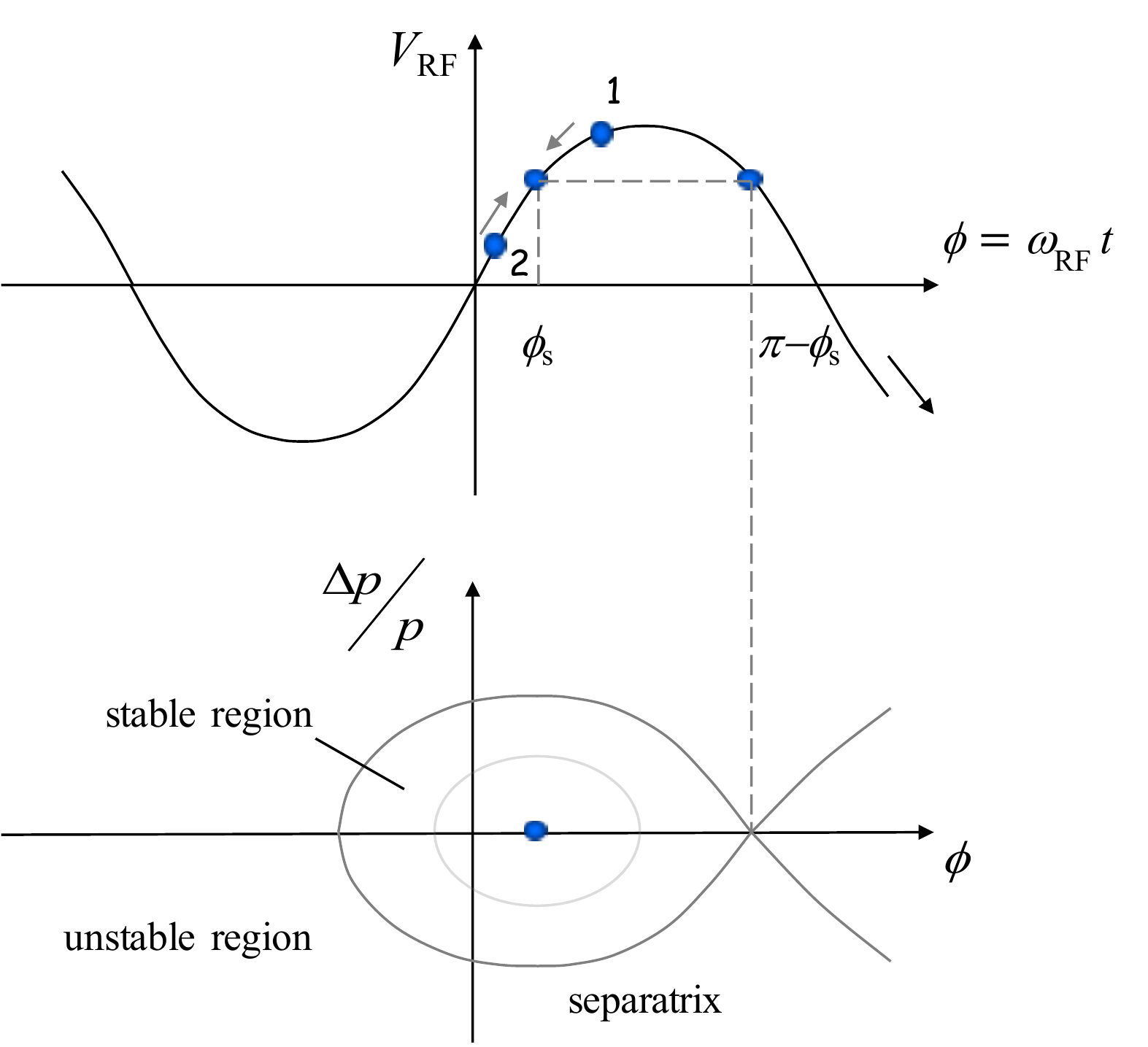}
 \caption{\label{fig:acc-bucket} Top graph shows RF voltage as a function of particle phase; bottom diagram is the phase-space picture}
\end{figure}

As we have seen earlier, the restoring force goes to zero when $\phi$ reaches $\pi - \phi_{\rm s}$, and it becomes non-restoring beyond (both below and above transition); see Fig.~\ref{fig:acc-bucket}.
Hence $\pi - \phi_{\rm s}$ is an extreme amplitude for a stable motion, which has a closed trajectory in phase space. This phase-space trajectory separates the region of stable motion from the unstable region; it is called the {\it separatrix}. The area within this separatrix is called the {\it RF bucket\/} and  corresponds to the maximum acceptance in phase space for a stable motion.

Since we have found an invariant of motion, we can write the equation for the separatrix by calculating it at the phase $\phi = \pi - \phi_{\rm s}$ where $\dot\phi = 0$:
\begin{equation}
\frac{\dot\phi^2}{2} - \frac{\Omega_{\rm s}^2}{\cos\phi_{\rm s}}  \left( \cos\phi + \phi \sin\phi_{\rm s} \right) = - \frac{\Omega_{\rm s}^2}{\cos\phi_{\rm s}}  \bigl( \cos(\pi -\phi_{\rm s}) + (\pi -\phi_{\rm s}) \sin\phi_{\rm s} \bigr) \: .
\label{eq:separatrix}
\end{equation}

From this we can calculate the second value, $\phi_{\rm m}$, where the separatrix crosses the horizontal axis, which is the other extreme phase for stable motion:
 \begin{equation}
 \cos\phi_{\rm m} + \phi_{\rm m} \sin\phi_{\rm s}  = \cos(\pi -\phi_{\rm s}) + (\pi -\phi_{\rm s}) \sin\phi_{\rm s} \, .
\end{equation}

It can be seen from the equation of motion that $\dot\phi$ reaches an extreme when $\ddot\phi = 0$, corresponding to $\phi = \phi_{\rm s}$. Putting this value into Eq. (\ref{eq:separatrix}) gives
\begin{equation}
\dot\phi^2_\mathrm{max} = 2 \,\Omega_{\rm s}^2 \bigl[ 2 + (2\phi_{\rm s}-\pi ) \tan\phi_{\rm s} \bigr] \: ,
\end{equation}
which translates into an acceptance in energy
\begin{equation}
\left( \frac{\Delta E}{E_{\rm s}} \right)_\mathrm{max} = \pm \beta \sqrt{ - \frac{e \hat{V}}{\pi h \eta E_{\rm s}} \,G(\phi_{\rm s}) } \; ,
\label{eq:energy_acc}
\end{equation}
where
\begin{equation}
G(\phi_{\rm s}) = 2 \cos\phi_{\rm s} + ( 2\phi_{\rm s}-\pi ) \sin\phi_{\rm s} \, .
\end{equation}

\begin{figure}[!hb]
 \centering\includegraphics[width=0.55\textwidth]{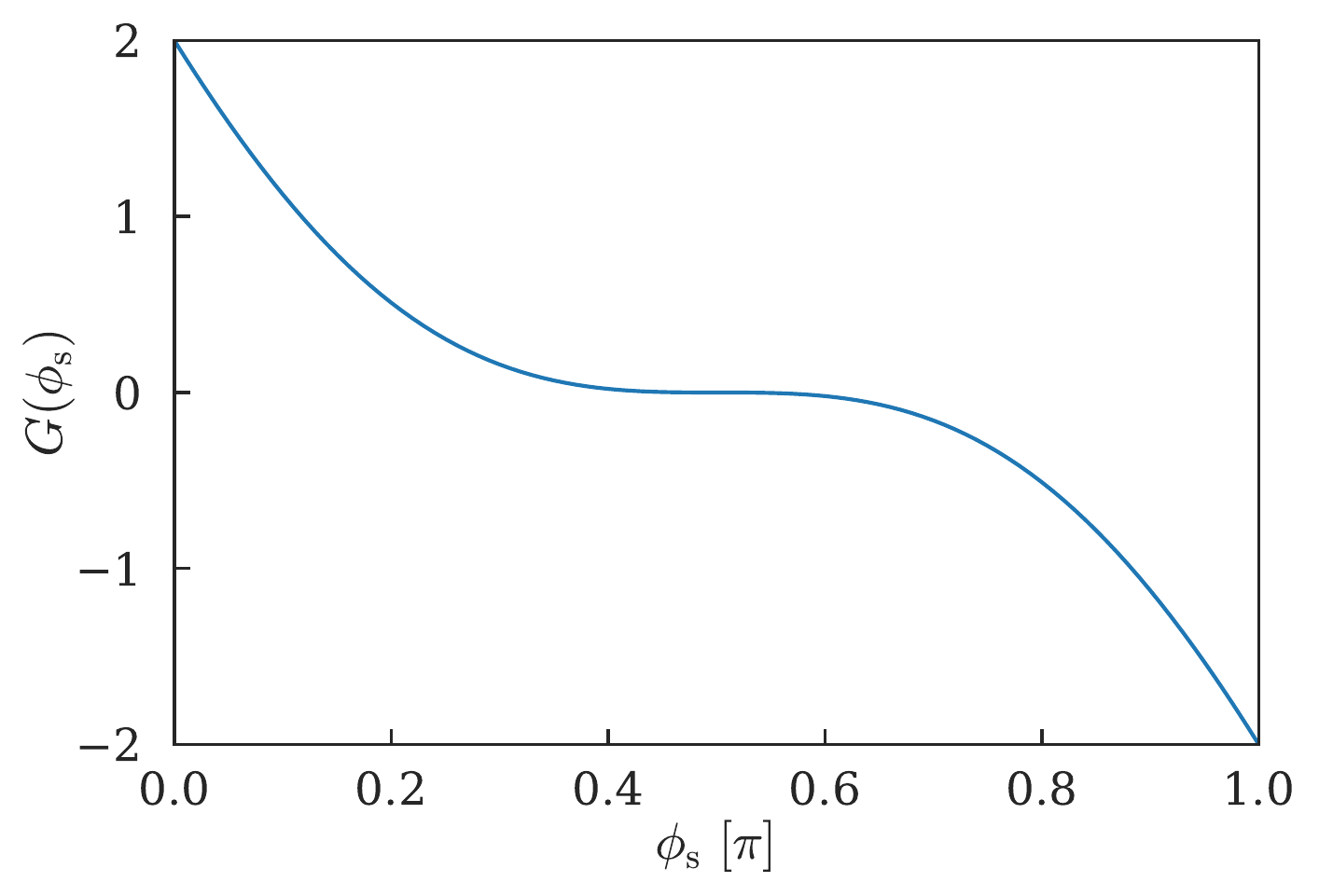}
 \caption{\label{fig:G} The function $G(\phi_{\rm s})$ for the energy acceptance in Eq.~(\ref{eq:energy_acc})}
\end{figure}

\begin{figure}[!hb]
 \centering\includegraphics[width=0.72\textwidth]{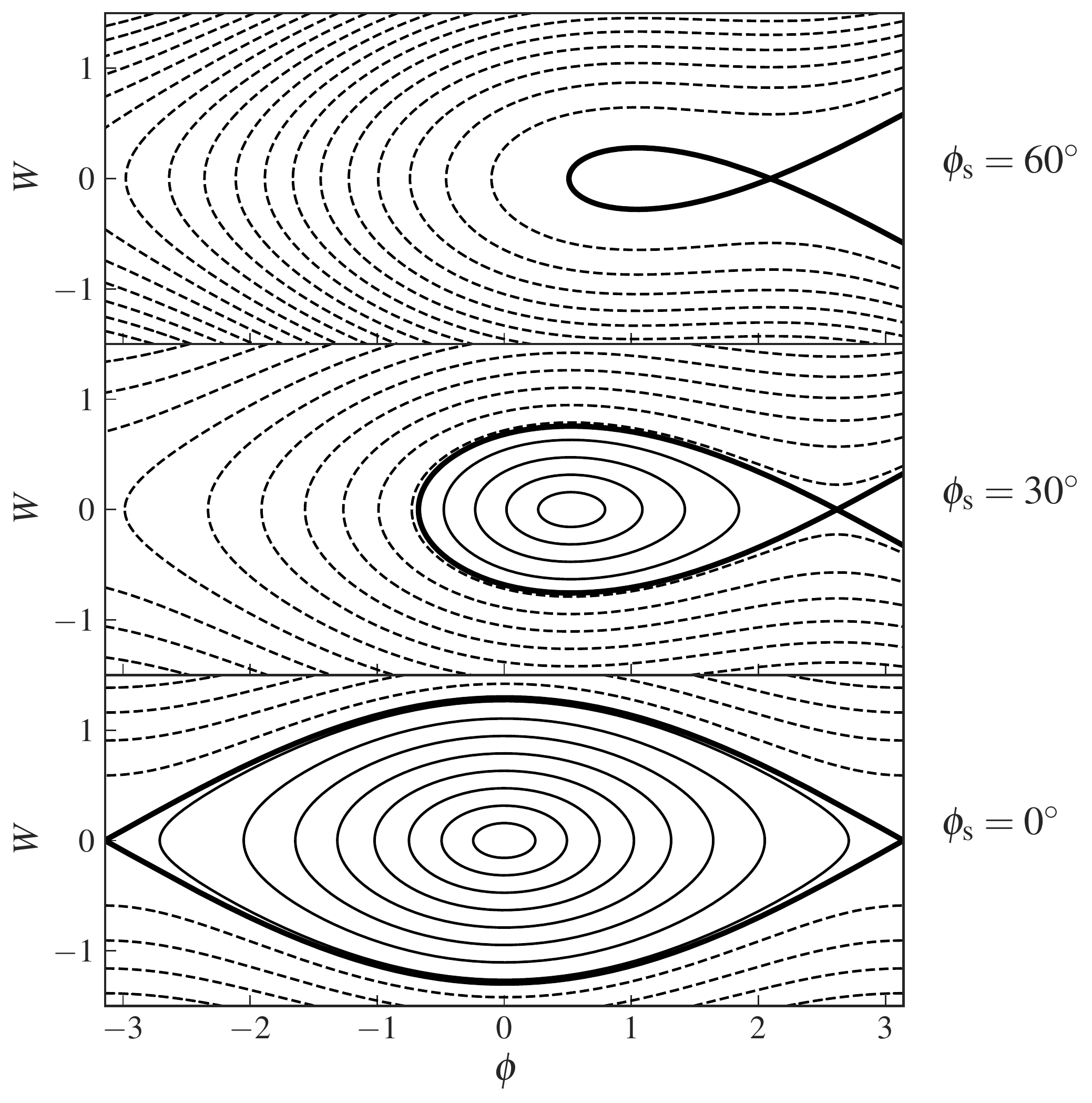}
 \caption{\label{fig:acceptance} Phase-space plots for different synchronous-phase angles $\phi_\mathrm{s}$ for otherwise identical parameters; in each plot, thin solid lines represent stable trajectories in phase space, dashed lines represent unstable trajectories, and the thick solid line is the separatrix.}
\end{figure}

This {\it RF acceptance} depends strongly on $\phi_{\rm s}$, as can be seen from the function $G(\phi_{\rm s})$ in Fig.~\ref{fig:G}, and plays an important role in the capture at injection and the stored beam lifetime.
The maximum energy acceptance in the bucket depends on the square root of the available RF voltage, $\hat{V}_{\rm RF}$.
The phase extension of the bucket is at a maximum for $\phi_{\rm s} = 0^\circ$ or $180^\circ$.
As the synchronous phase approaches $90^\circ$, the bucket size becomes smaller, as illustrated in Fig.~\ref{fig:acceptance}.

\subsection{Stationary bucket}
In the case of the stationary bucket, we have no acceleration and $\sin\phi_{\rm s} = 0$, so  $\phi_{\rm s} = 0$ or $\pi$. The equation of the separatrix for $\phi_{\rm s} = \pi$ (above transition) simplifies to
\begin{equation}
\frac{\dot\phi^2}{2} + \Omega_{\rm s}^2 \cos\phi = \Omega_{\rm s}^2 \qquad \mathrm{or} \qquad \frac{\dot\phi^2}{2} = 2 \, \Omega_{\rm s}^2 \sin^2\frac{\phi}{2}
\; .
\end{equation}

At this point, it is convenient to introduce a new variable $W$ to replace the phase derivative $\dot\phi$:
\begin{equation}
W = \frac{\Delta E}{\omega_{\rm rs}} = - \frac{p_{\rm s} R_{\rm s}}{h \eta \omega_{\rm rs}} \,\dot\phi \, ,
\end{equation}
where $\omega_{\rm rs}$ is the revolution frequency of the synchronous particle. As we shall see later, this new variable is canonical. Different choices of canonical variables are possible and lead to slightly different equations, as, for example, in Ref.~\cite{bib:CAS2013-tecker}.

Introducing $\Omega_{\rm s}^2$ from Eq. (\ref{eq:osc}) leads to the following equation for the separatrix:
\begin{equation}
W = \pm \frac{C}{\pi c} \sqrt{ \frac{-e \hat{V} E_{\rm s}}{2 \pi h \eta}}\:  \sin\frac{\phi}{2} = \pm W_{\rm bk} \sin\frac{\phi}{2} \qquad \text{with } \;
W_{\rm bk} = \frac{C}{\pi c} \sqrt{ \frac{-e \hat{V} E_{\rm s}}{2 \pi h \eta}}
\; .
\end{equation}

Setting $\phi = \pi$ in the previous equation shows that $W_{\rm bk}$ is the maximum height of the bucket, which results in the maximum energy acceptance
\begin{equation}
\Delta E_\mathrm{max} = \omega_{\rm rs} W_{\rm bk} = \beta_{\rm s} \sqrt{ 2 \frac{-e \hat{V}_{\rm RF} E_{\rm s}}{\pi h \eta}} \; .
\end{equation}

The bucket area is
\begin{equation}
A_{\rm bk} = 2 \int_0^{2\pi} W \,\mathrm{d}\phi \: .
\end{equation}
With $\int_0^{2\pi} \sin(\phi/2) \, \mathrm{d}\phi = 4$, one obtains
\begin{equation}
A_{\rm bk} = 8 W_{\rm bk} = 8 \frac{C}{\pi c} \sqrt{ \frac{-e \hat{V} E_{\rm s}}{2 \pi h \eta}} \; .
\end{equation}

\subsection{Bunch matching into the stationary bucket}
\begin{figure}[!htb]
 \centering\includegraphics[width=0.48\textwidth]{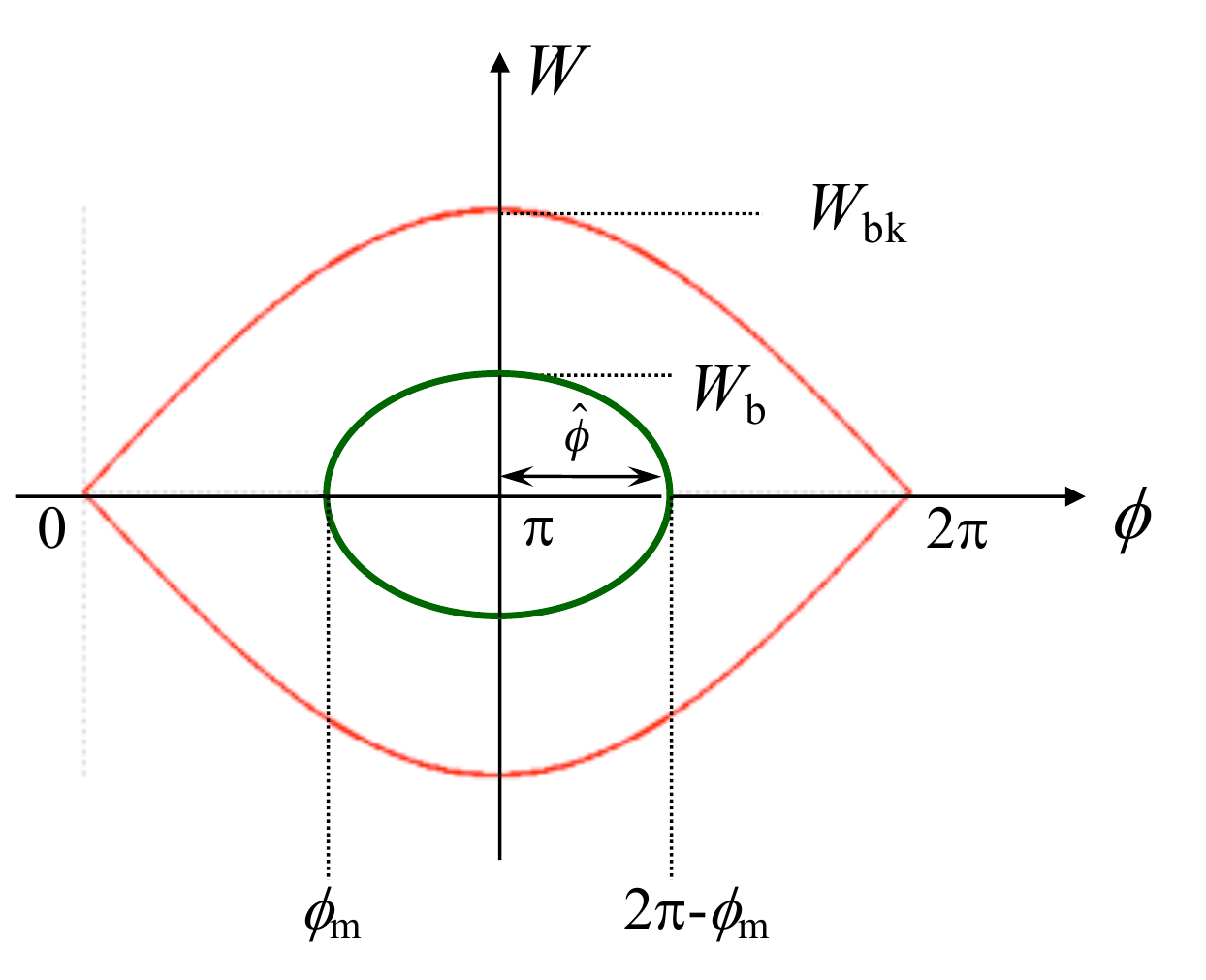}
 \caption{\label{fig:stationary-bucket} Phase-space plot for the separatrix of the stationary bucket and a trajectory inside}
\end{figure}

We can describe the motion of a particle inside the separatrix of the stationary bucket
by starting from the invariant of motion in Eq. (\ref{eq:invariant}) and setting $\phi_{\rm s} = \pi$:
\begin{equation}
\frac{\dot\phi^2}{2} + \Omega_{\rm s}^2 \,\cos\phi = I \, .
\end{equation}
The points $\phi_{\rm m}$ and $2\pi-\phi_{\rm m}$ where the trajectory crosses the horizontal axis are symmetrical with respect to $\phi_{\rm s} = \pi$ (see Fig. \ref{fig:stationary-bucket}). 
We can calculate the invariant for $\phi = \phi_{\rm m}$ and get
\begin{equation}
\frac{\dot\phi^2}{2} + \Omega_{\rm s}^2 \,\cos\phi = \Omega_{\rm s}^2 \,\cos\phi_{\rm m}\:,
\end{equation}
\begin{equation}
\dot\phi = \pm \Omega_{\rm s} \sqrt{2\, (\cos\phi_{\rm m} - \cos\phi)}\; ,
\end{equation}
\begin{equation}
W = \pm W_{\rm bk} \sqrt{\cos^2\frac{\phi_{\rm m}}{2} - \cos^2\frac{\phi}{2}} \qquad \Bigl( \mathrm{using } \; \cos\phi = 2 \cos^2\frac{\phi}{2} - 1 \Bigr)\; .
\end{equation}

Setting $\phi = \pi$ in the previous equation allows us to calculate the bunch height $W_{\rm b}$:
\begin{equation}
W_{\rm b} = W_{\rm bk} \,\cos\frac{\phi_{\rm m}}{2} = W_{\rm bk} \,\sin\frac{\hat\phi}{2}
\: ,
\end{equation}
with $\hat\phi = \pi - \phi_{\rm m}$ being the maximum phase amplitude for an oscillation around the synchronous phase $\phi_{\rm s} = \pi$.

The corresponding maximum energy difference of a particle on this phase-space trajectory is
\begin{equation}
\left( \frac{\Delta E}{E_{\rm s}} \right)_{\!{\rm b}} = \left( \frac{\Delta E}{E_{\rm s}} \right)_{\rm RF} \cos\frac{\phi_{\rm m}}{2} = \left( \frac{\Delta E}{E_{\rm s}} \right)_{\rm RF} \sin\frac{\hat\phi}{2} \; .
\end{equation}

When a particle bunch is injected into a synchrotron, the bunch has a given bunch length and energy spread. Each of the different particles will move along a phase-space trajectory that corresponds to its initial phase and energy. If the shape of the injected bunch in phase space matches the shape of a phase-space trajectory for the given RF parameters, the shape of the bunch in phase space will be maintained.

\begin{figure}[!b]
 \centering\includegraphics[width=0.95\textwidth]{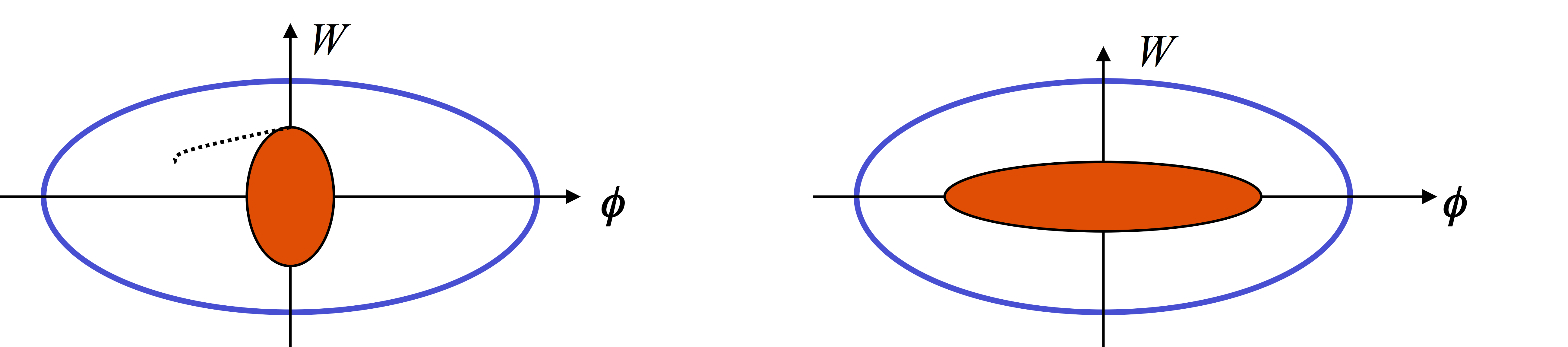}
 \caption{\label{fig:matching} Phase-space plots for a mismatched bunch one-quarter of a synchrotron period apart}
\end{figure}

If the shape does not match, it will vary during the synchrotron period. This is illustrated in Fig.~\ref{fig:matching} for a bunch that has a shorter bunch length and a larger energy spread compared with the phase-space trajectory. As the particles move along their individual trajectories, the bunch will be longer with a smaller energy spread after one-quarter of a synchrotron period, and it will regain the initial shape after one-half of a period. This effect can be used to manipulate the shape of the bunch in phase space and trade off bunch length against energy spread (so-called {\it bunch rotation}). When the RF voltage in matched conditions is suddenly increased, the bunch will be shorter after a quarter of a synchrotron period. In this way, it can be shortened for a transfer to a higher-frequency RF system.

\begin{figure}[!hb]
 \centering\includegraphics[width=0.97\textwidth]{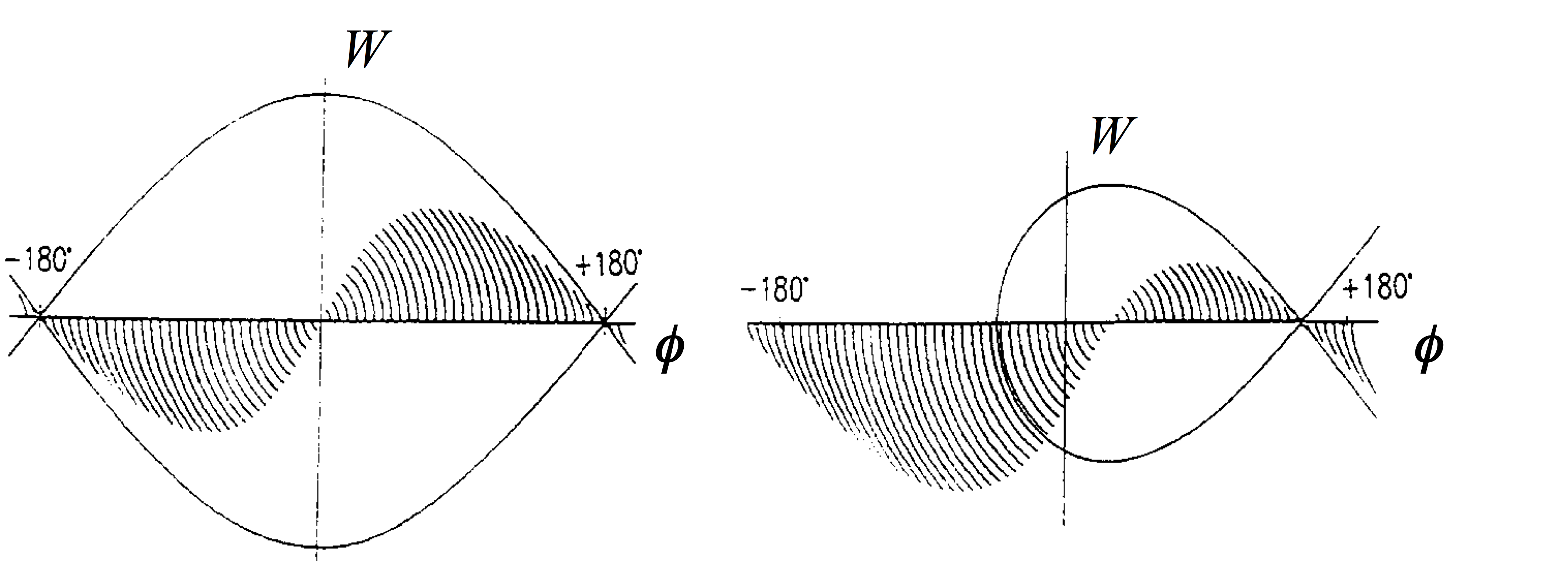}
 \caption{\label{fig:bucket-rotation} Phase-space trajectories for one-eighth of the synchrotron period: left, stationary bucket; right, accelerating bucket \cite{pirkl}.}
\end{figure}

Owing to the non-linear restoring force, the synchrotron period depends on the oscillation amplitude, and particles with larger amplitudes have a longer synchrotron period, as shown in Fig. \ref{fig:bucket-rotation}. This will eventually lead to {\it filamentation} of the bunch and an increase in  the longitudinal emittance.

The same phenomenon will occur when the bunch shape is matched to the bucket but there is an error in the phase or the energy. The different particles in the bunch
will perform their individual oscillations around the synchronous particle and will filament, leading to an increase in the longitudinal emittance. To avoid an emittance increase, it is important to match the phase, the energy, and the shape of the bunch and the bucket during the transfer.

\subsection{Potential energy function and Hamiltonian}
The longitudinal motion is produced by a force that can be derived from a scalar potential $U$:
\begin{align}
\der{^2\phi}{t^2} &= F(\phi) = - \frac{\partial U}{\partial\phi} \: ,
\\
U(\phi) &= - \int\limits_0^\phi F(\phi) \, \rd\phi = - \frac{\Omega^2_{\rm s}}{\cos\phi_{\rm s}} \,( \cos\phi + \phi\sin\phi_{\rm s} ) - F_0
\: .
\end{align}
The sum of the potential energy and the kinetic energy is constant and, by analogy, represents the total energy of a non-dissipative system:
\begin{equation}
\frac{\dot\phi^2}{2} + U(\phi) = F_0 \, .
\end{equation}

Since the total energy is conserved, we can describe the system as a Hamiltonian system. Different choices of the canonical variables are possible.
With the variable
\begin{equation}
W = \frac{\Delta E}{\omega_{\rm rs}} \, ,
\end{equation}
the two first-order equations of the longitudinal motion become
\begin{align}
\der{\phi}{t} & =  - \frac{h \eta \omega_{\rm rs}}{p R} \,W \label{eq:dphi}
\, , \\
\der{W}{t} & =  \frac{e \hat{V}}{2\pi} \, ( \sin\phi - \sin\phi_{\rm s} ) \: .
\end{align}

The two variables $\phi$ and $W$ are canonical, since these equations of motion can be derived from a Hamiltonian $H(\phi, W, t)$:
\begin{align}
\der{\phi}{t} &= \pder{H}{W}\: , \qquad\der{W}{t} = - \pder{H}{\phi} \: ,
\\
H(\phi,W, t) &= \frac{e \hat{V}}{2\pi} \big[ \cos\phi - \cos\phi_{\rm s} + (\phi-\phi_{\rm s}) \sin\phi_{\rm s} \big] - \frac{1}{2} \frac{h \eta \omega_{\rm rs}}{p R} W^2 \, .
\label{eq:ham}
\end{align}
The basic Hamiltonian shown here reproduces the equations of motion that we found previously.
In more complex cases, the general approach of the Hamiltonian formalism can help us to analyse and understand some fairly
complicated dynamics (multiple harmonics, bunch splitting, \etc).

\begin{figure}[!h]
\begin{minipage}{\textwidth}
  \centering
  $\vcenter{\hbox{\includegraphics[width=0.59\textwidth]{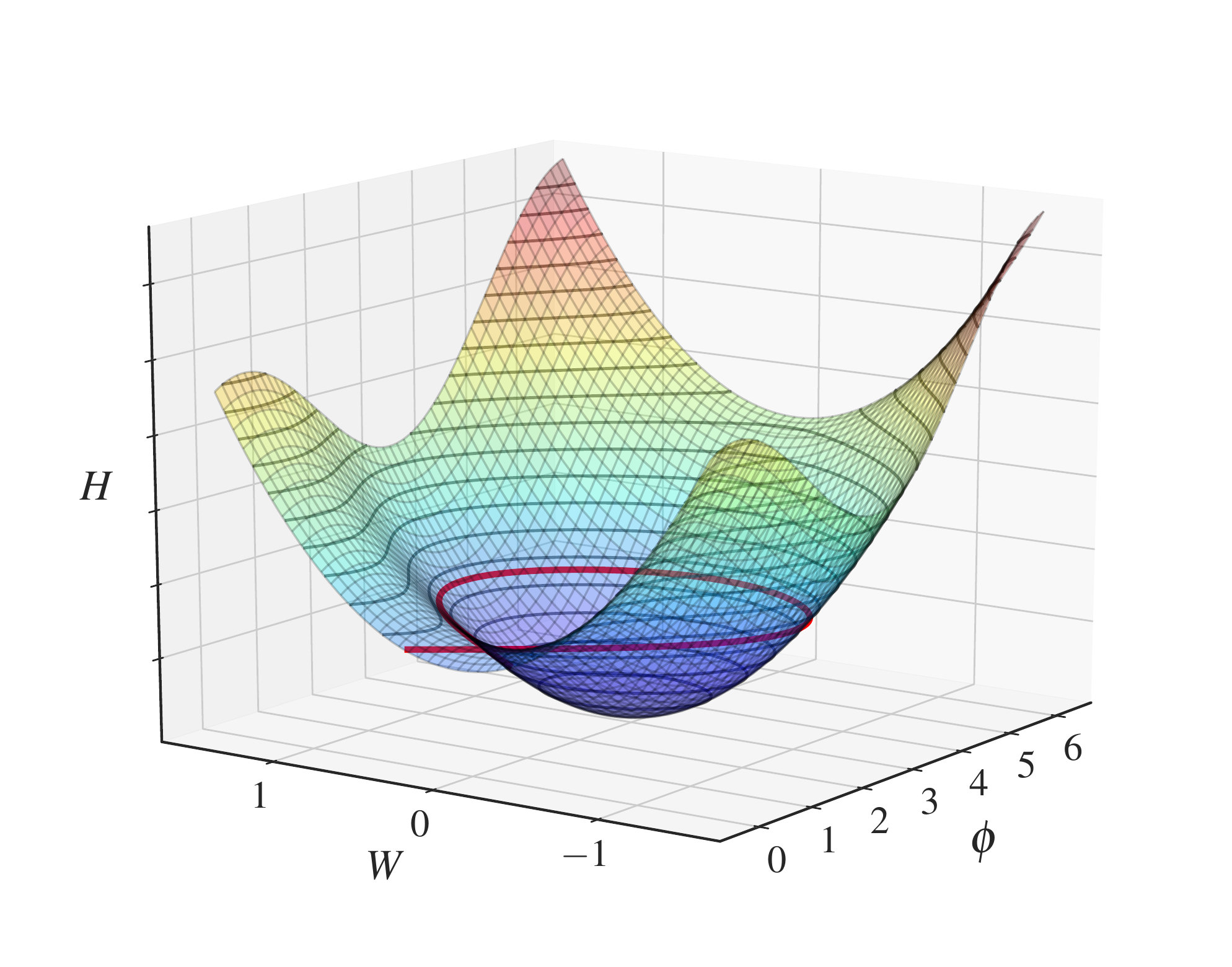}}}$
  \hfill
  $\vcenter{\hbox{\includegraphics[width=0.4\textwidth]{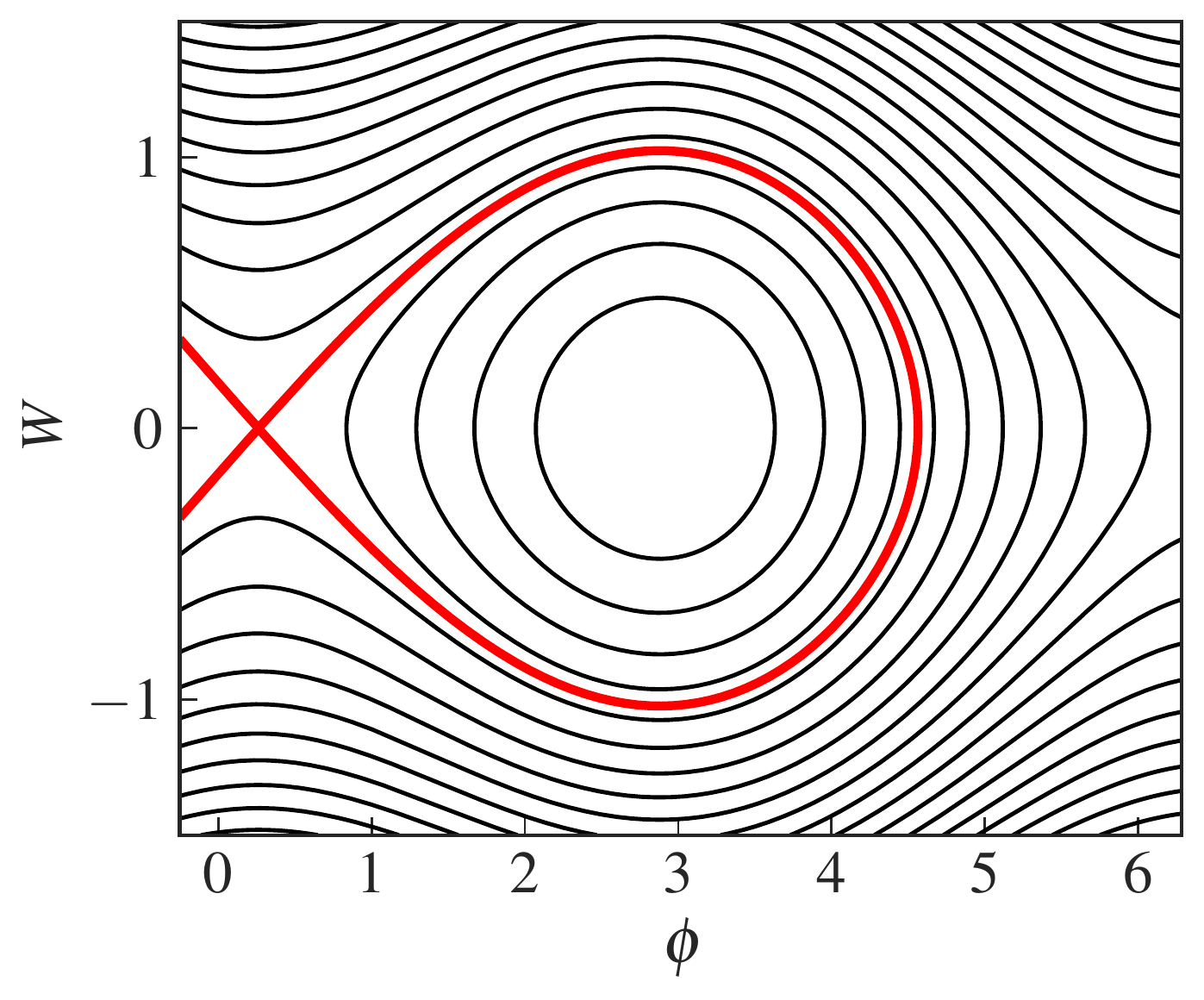}}}$
\end{minipage}
 \caption{\label{fig:phase-space-3d} Example of  a Hamiltonian function $H$ (left) and its projection (right) above transition, for a synchronous phase of $\phi_s=170^\circ$. Equipotential lines are possible phase-space trajectories; the separatrix is shown as a solid red  line.}
\end{figure}

The Hamiltonian $H$ represents the total energy of the system. In fact, if the total energy is conserved,
the contours of constant $H$ are particle trajectories in phase space, as illustrated in Fig.~\ref{fig:phase-space-3d}.

\section*{Acknowledgement}

I thank, in particular, Jo\"el Le Duff, who gave this lecture series in the past and allowed me to use the material of his lectures, from which I profited greatly. Many of the figures and derivations in this paper originate from him.

\begin{flushleft}

\section*{Bibliography}
J. Le Duff, Proc.\ CAS-CERN Accelerator School: CAS Fifth General Accelerator Physics Course, Jyv\"{a}skyl\"{a}, Finland, 7--18 September 1992 (CERN 94-01). Ed.\ S.\ Turner  (CERN, Geneva, 1994), pp.~253--288, \url{https://doi.org/10.5170/CERN-1994-001.253}.

\noindent J. Le Duff, Proc.\ CAS-CERN Accelerator School: CAS Fifth General Accelerator Physics Course, Jyv\"{a}skyl\"{a}, Finland, 7--18 September 1992 (CERN 94-01). Ed.\ S.\ Turner (CERN, Geneva, 1994), pp.~289--311, \url{https://doi.org/10.5170/CERN-1994-001.289}.

\noindent H. Wiedemann, \textit{Particle Accelerator Physics} (Springer, Berlin, 2007),\\ \url{https://doi.org/10.1007/978-3-540-49045-6}.

\noindent K. Wille, \textit{The Physics of Particle Accelerators: An Introduction}\\ (Oxford University Press, Oxford, 2000).

\noindent T. Wangler, \textit{RF Linear Accelerators} (Wiley-VCH, Weinheim, 2008),\\ \url{https://doi.org/10.1002/9783527623426}.

\end{flushleft}


\begin{thebibliography}{99}

\bibitem{wilson}
E.J.N. Wilson, \textit{An Introduction to Particle Accelerators} (Oxford University Press, Oxford, 2001), \url{https://doi.org/10.1093/acprof:oso/9780198508298.001.0001}.

\bibitem{bib:CAS2013-tecker}
F. Tecker, Proc.\ CAS-CERN Accelerator School: Advanced Accelerator Physics, Trondheim, Norway, 19--29 August 2013 (CERN-2014-009). Ed.\ W. Herr  (CERN, Geneva, 2014), pp.~1--21, \url{https://doi.org/10.5170/CERN-2014-009.1}.

\bibitem{pirkl}
W. Pirkl, Proc.\ CAS-CERN Accelerator School: CAS Fifth Advanced Accelerator Physics Course, Rhodes, Greece, 20 September--1 October 1993 (CERN 95-06). Ed.\ S.\ Turner (CERN, Geneva, 1995), pp.~233--257, \url{https://doi.org/10.5170/CERN-1995-006.233}.
\end{thebibliography}
\end{document}